\documentclass[conference]{IEEEtran}
\IEEEoverridecommandlockouts
\usepackage{cite}
\usepackage{amsmath,amssymb,amsfonts}
\usepackage{algorithmic}
\usepackage{graphicx}
\usepackage{textcomp}
\usepackage{xcolor}
\usepackage{multirow}
\usepackage{listings}
\usepackage{adjustbox}
\usepackage{float}
\usepackage{pifont}
\def\BibTeX{{\rm B\kern-.05em{\sc i\kern-.025em b}\kern-.08em
    T\kern-.1667em\lower.7ex\hbox{E}\kern-.125emX}}

\def\code#1{\texttt{#1}}
\begin{document}

\title{Mining Function Homology of Bot Loaders from Honeypot Logs
}
\author{Yuhui Zhu, Zhenxiang Chen, Qiben Yan, Shanshan Wang, Enlong Li, Lizhi Peng, Chuan Zhao}

\maketitle

\begin{abstract}
Self-contained loaders are widely adopted in botnets for injecting loading commands and spawning new bots. 
While researchers can dissect bot clients to get various information of botnets, the cloud-based and self-contained design of loaders effectively hinders researchers from understanding the loaders' evolution and variation using classic methods.
The decoupled nature of bot loaders also dramatically reduces the feasibility of investigating relationships among clients and infrastructures.
In this paper, we propose a text-based method to investigate and analyze details of bot loaders using honeypots.
We leverage high interaction honeypots to collect request logs and define eight families of bot loaders based on the result of agglomerative clustering.
At the function level, we push our study further to explore their homological relationship based on similarity analysis of request logs using sequence aligning techniques.
This further exploration discloses that the released code of Mirai keeps spawning new generations of botnets both on the client and the server side.
This paper uncovers the homology of active botnet infrastructures, providing a new prospect on finding covert relationships among cybercrimes.
Bot loaders are precisely investigated at the function level to yield a new insight for researchers to identify the botnet's infrastructures and track their evolution over time.

\end{abstract}


\section{Introduction}
Accompanying the growth of the IoT market, botnets targeting IoT devices have become a major threat to cyber security.
The IoT botnet is a worm-like malware system targeting monopolizing IoT devices to execute malicious actions.
Nowadays, IoT Botnets have become the most popular solution for attackers conducting DDoS attacks\cite{koliasDDoSIoTMirai2017}, exploiting other hosts\cite{dangUnderstandingFilelessAttacks2019} and even cryptomining\cite{bijmansJustTipIceberg2019}.
Mirai\cite{antonakakisUnderstandingMiraiBotnet2017,malwaremustdie!MMD00562016LinuxMirai2016}, the most famous descendant of classic botnet Bashlite, was recorded launching a 1.1Tbps DDoS attack using 148,000 IoT devices, which broke the record and makes it the top botnet in the following years.
This large family has been constantly active since its inception.
Anna-senpai's code release\cite{anna-senpaiFREEWorldLargest2016} in September 2016 further stimulated botmasters to create variants based on its idea.
New botnets like Hajime\cite{herwigMeasurementAnalysisHajime2019} are also under active development to equip themselves with new technologies.

Security industries leverage honeypots\cite{fanEnablingAnatomicView2018} to collect comprehensive information on botnet threats, including malware samples, interaction logs, and attackers' IP addresses, then publish Indicators of Compromise (IoC) as their product or service.
To better understand botnets, some studies\cite{cozziTangledGenealogyIoT2020,jangAutomaticSoftwareLineage2013,dibMultiDimensionalDeepLearning2021,ghietteFingerprintingToolingUsed2019} introduced malware lineage inference into this field for a systematic insight into the relationships of botnets.
However, due to the decoupled design of popular IoT botnets, honeypots cannot collect every component of botnets, especially executables of bot loaders.
Most botnets use a self-contained loader server to drop bot executables during the propagating period, which executable can never be captured by honeypots.
The limited dataset forces systematic studies focusing on dissecting bot executables to investigate the relationship among botnets.
Bot loaders are less investigated, which dramatically reduces the feasibility of inferring the infrastructures' evolution and transfers conducted by botmasters.

Despite it is impractical to dissect bot loaders' executables, some studies \cite{torabiStringsBasedSimilarityAnalysis2021,tabariWhatAreAttackers2021,lingenfelterAnalyzingVariationIoT2020,antonakakisUnderstandingMiraiBotnet2017,griffioenExaminingMiraiBattle2020,stephensDetectingInternetThings2021} made elementary comparative analyses on specified properties in telnet interaction, including initialize command lists, credential dictionaries, and query tokens.
These studies reached a preliminary conclusion that the relationships among bot loaders can be inferred from their shared features, but did not make any comparison on their intrusion behaviors.
Though the homology of bot loaders' functions is critical for inferencing their covert relationships, the lack of methodology hinders researchers from understanding bot loaders and their codebases on a systematic aspect.

The main purpose of this paper is to reveal the homology of active botnets' infrastructures, precisely speaking, investigating bot loaders on the function level to find their undercovered homology relationship.
To achieve this goal, we apply NLP-based methods to classify interaction logs captured by telnet honeypots and extract stable strings from interaction logs to represent the behavior feature of every bot loader family.
We provide a new homology view on bot loaders, helping researchers obtain deeper knowledge about IoT botnets.
The main contributions of our work are summarized as follows:

\begin{itemize}
  \item
    Our work provides a server-side perspective for cyber forensics to investigate cybercrimes beside bots.
    We try to describe loaders' functions using request logs captured by the telnet honeypot, which is an alternative method to understand the inner details of bot loaders.
  \item
    We propose a semantic-aware yet an elastic method for classifying bot loaders based on interaction logs from honeypots.
    We define 8 families out of 19 clusters and describe their behavior similarity based on the result of agglomerative clustering and several empirical criteria.
    We also extracted representative strings for each cluster to show their functional characters, which provides an explainable result for our family definition.
  \item 
    We made an in-depth investigation of loaders' behaviors and their relationships.
    To examine the component homologies of loaders, we analyze extracted representative strings associated with the result of agglomerative clustering.
    The common skeleton and shared components across different families are well described based on these results.
    This prospect can provide solid evidence for investigating underground industries to uncover their connections.
\end{itemize}

The rest of the paper is organized as follows:
In Section \ref{sec_background}, we introduce the basics of IoT botnets and the status of botnet studies. We also discuss the limitations of current studies on botnet infrastructures.
In Section \ref{sec_methodology}, we introduce our honeypot system and propose our classification method. Based on self-supervised learning, we classify request logs and provide a primary classification of active loaders.
In Section \ref{sec_eval}, we evaluate the result in Section \ref{sec_methodology} to introduce our definition of loader families. We describe their functions and depict their homology based on their common behaviors.
Finally, we discuss the limitation of our work in Section \ref{sec_discussion} and conclude our work in Section \ref{sec_conclusion}.

\section{Related Works}
\label{sec_background}
\subsection{Basics on IoT Botnets}

On the dark side, since the industrial community introduced ``Internet of Things", cybercrimes had targeted them and developed various botnet software to exploit vulnerabilities on these platforms and devices.

\begin{figure}[tp]
  \centering
  \includegraphics[width=0.8\linewidth]{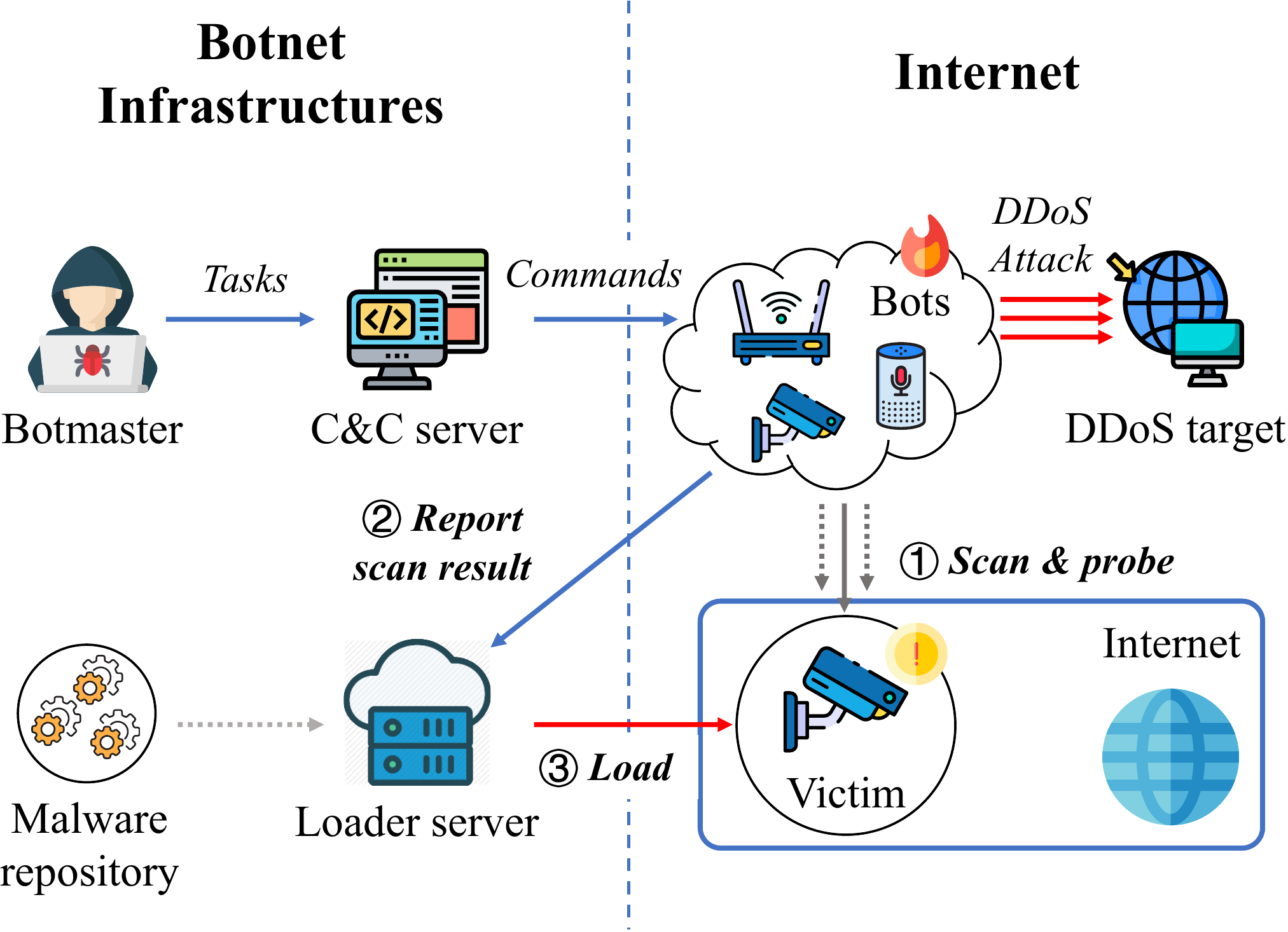}
  \caption{Common botnet architecture.}
  \label{fig_botnet_arch}
\end{figure}

In the Anna-senpai's release\cite{anna-senpaiFREEWorldLargest2016}, the botnet was designed as a distributed system that consists of 3 major components, \emph{bot}, \emph{loader server}, and \emph{C\&C server}.
The overall architecture is dipicted in Fig. \ref{fig_botnet_arch}.
In the attack phase (\ding{172}), the C\&C server sends attack commands to bots like traditional botnets.
But in the propagate phase, bots only scan for vulnerable devices but never load bot executables (\ding{173}).
They merely report collected IP addresses and credentials to the loader server.
The self-contained loader server executes loading scripts to launch bot malware on victim devices (\ding{174}).
The decoupled loader and distributed architecture simplified its design and robustified its infrastructure to survive taking-downs.

\subsection{Investigating and Tracing Botnets}

On the light side, the dynamics of botnets are traced closely to investigate their background information.
Firstly, researchers collect everything about how botnets work and propagate, which is the fundamental identity of malicious activities.
Secondly, they make a comprehensive inspection of threats to investigate the criminal on the dark side.
Lastly, they catch up with the dynamic of these botnets to keep a long-term track of these botnets.

\begin{figure*}[htb]
  \centering
  \includegraphics[width=\textwidth]{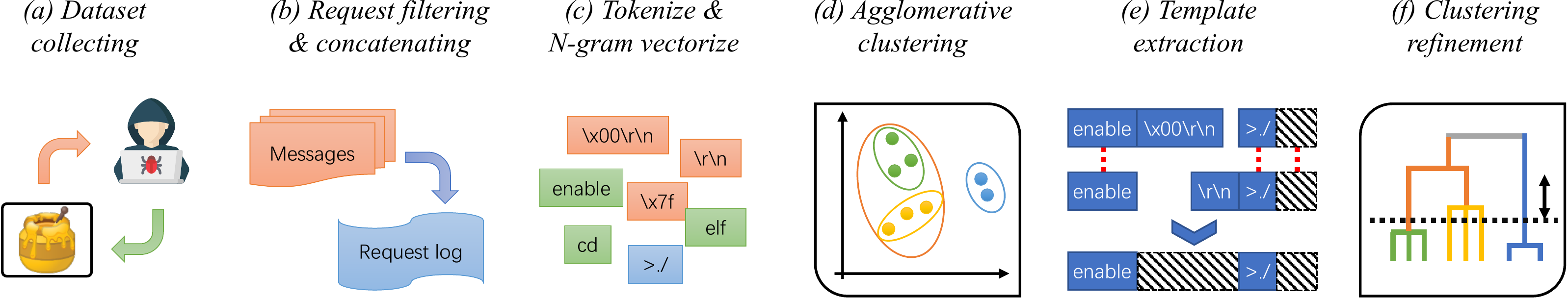}
  \caption{Schematic depiction of the analysis steps in our work.}
  \label{fig_methodology}
\end{figure*}

Honeypot is one of the most commonly used tools for collecting threat information, especially botnets' dynamics.
As defined by Spitzner\cite{spitznerHoneypotsCatchingInsider2003}, Honeypot is ``an information system resource whose value lies in the unauthorized or illicit use of that resource which acts like a vulnerable device to interact with malicious adversaries''.
Honeypots record intrusion conversations from attackers, including the IP address and network interactions.
Sometimes it also stores loaded executables for further analyses.
Pa et al.\cite{paIoTPOTAnalysingRise2015} first introduced honeypot into the IoT domain to capture botnet malwares.

To monitor botnets and observe their evolution, researchers continuously gather information and track dynamics to follow up on botnets' entire lifecycle.
Emerging botnets and infected devices are continuously measured to summarize the dynamics of threats.
The first research about Mirai made by Antonakakis et al.\cite{antonakakisUnderstandingMiraiBotnet2017} tried to observe Mirai's spread and evolution from many aspects.
This work utilized a network telescope, scanner, honeypot, DNS probe, and C\&C milker to collect information, then analyzed DDoS records provided by CDN providers.
In this paper, the authors recorded the growth and evolution ever since Mirai first appears, then summarized details about infections, ownerships, and attacks.
By relating C\&C servers, C\&C domain names, and credential dictionaries, this work examined several individual botnets and observed the expansion of their variants.
The Hajime botnet was analyzed in detail by \cite{herwigMeasurementAnalysisHajime2019}.
This work actively tracks bot nodes of Hajime on the DHT network.
This work also investigated injection requests based on the DNS backscatter dataset.
Griffioen's work\cite{griffioenExaminingMiraiBattle2020} took a wider view of all descendants of Mirai.
This work tracked how these botnets infect, retain control, and hijack devices from other botnets using virtual high-interaction honeypots, then demonstrated the mutation and evolution of Mirai in these years.
Wang et al. \cite{wangEvolutionaryStudyIoT2021} applied an NLP-based method to correlate the information published by researchers and constructed a lineage graph for 72 IoT malware families to reflect their evolutionary relationships.

To keep up with the evolving network environment, botnets also evolved to enhance themselves in exploitation by integrating new technologies.
Tracking emerging exploitations is a fundamental issue for tracking botnets' evolution and helping IDS systems protect devices.
Pa's work\cite{paIoTPOTAnalysingRise2015} made an elementary analysis based on credential dictionaries, intrusion commands, and hash values of executable samples.
Dang et al.\cite{dangUnderstandingFilelessAttacks2019} focused on file-less attacks conducted by IoT botnets and classified common types using a hybrid honeypot system.
Safaei et al.\cite{safaeipourDatadrivenCurationLearning2020} proposed a method aiming at classifying and inferring compromised IoT devices by solely observing one-way network traffic. This helps uncover, report, and thoroughly analyze ``in the wild'' IoT botnets on the Internet scale.
Abbas et al.\cite{abbasGenericSignatureDevelopment2021} made a comparative analysis of Mirai and Bashlite variants by analyzing collected attack vectors and malware executables.

\subsection{Comparative Study on Bot Loaders}

The released source code inspired botmasters to engage new techniques in the existing version.
This also boosted the diversity of bot loaders.
Comparing behavior features of loaders become a significant topic for understanding the evolution of botnets.
Lingenfelter et al.\cite{lingenfelterAnalyzingVariationIoT2020} analyzed interaction logs from medium interaction honeypots, made a comparison of initial commands and query tokens to demonstrate the variation among IoT botnets.
Torabi et al.\cite{torabiStringsBasedSimilarityAnalysis2021} tried mining unique strings from logs to build associations among active botnets.
Tabari et al.\cite{tabariWhatAreAttackers2021} made a statical analysis of the most commonly exploited vulnerabilities, credentials, and intrusion commands.

Although these works introduced a comparison aspect on active bot loaders, their conclusions are still less solid to reach a systematic conclusion on how the loaders variated and derived predecessors' codebase.
Some studies\cite{antonakakisUnderstandingMiraiBotnet2017,griffioenExaminingMiraiBattle2020}  adopted password dictionaries to track botnets' variation and analyze their lineage, but these works didn't focus on the implementation of intrusion commands, resulting in insufficient evidence for homology analysis of loaders.

\subsection{Summary}

The existing research mostly focused on revealing botnets' behavior in specified fields.
Lineage analyses are commonly applied on executables of bot clients, but infrastructures, especially loaders, are less investigated due to their self-contained design and cloud-side deployment.
Analyzing these fundamental components of botnets needs better mining and representing methods for monitored exploits.
This hinders researchers from extracting their implementation details, and also makes it difficult to analyze their evolution and homology.

\section{Methodology}
\label{sec_methodology}
In this section, we propose a novel methodology for investigating the relationships of bot loaders by analyzing their homology.
Based on the agglomerated clusters of the dataset, we extract the command-based features of loader families and try to figure out the lineage among different loaders.

\subsection{Dataset Collecting}

\begin{figure}[bp]
  \centering
  \includegraphics[width=\linewidth]{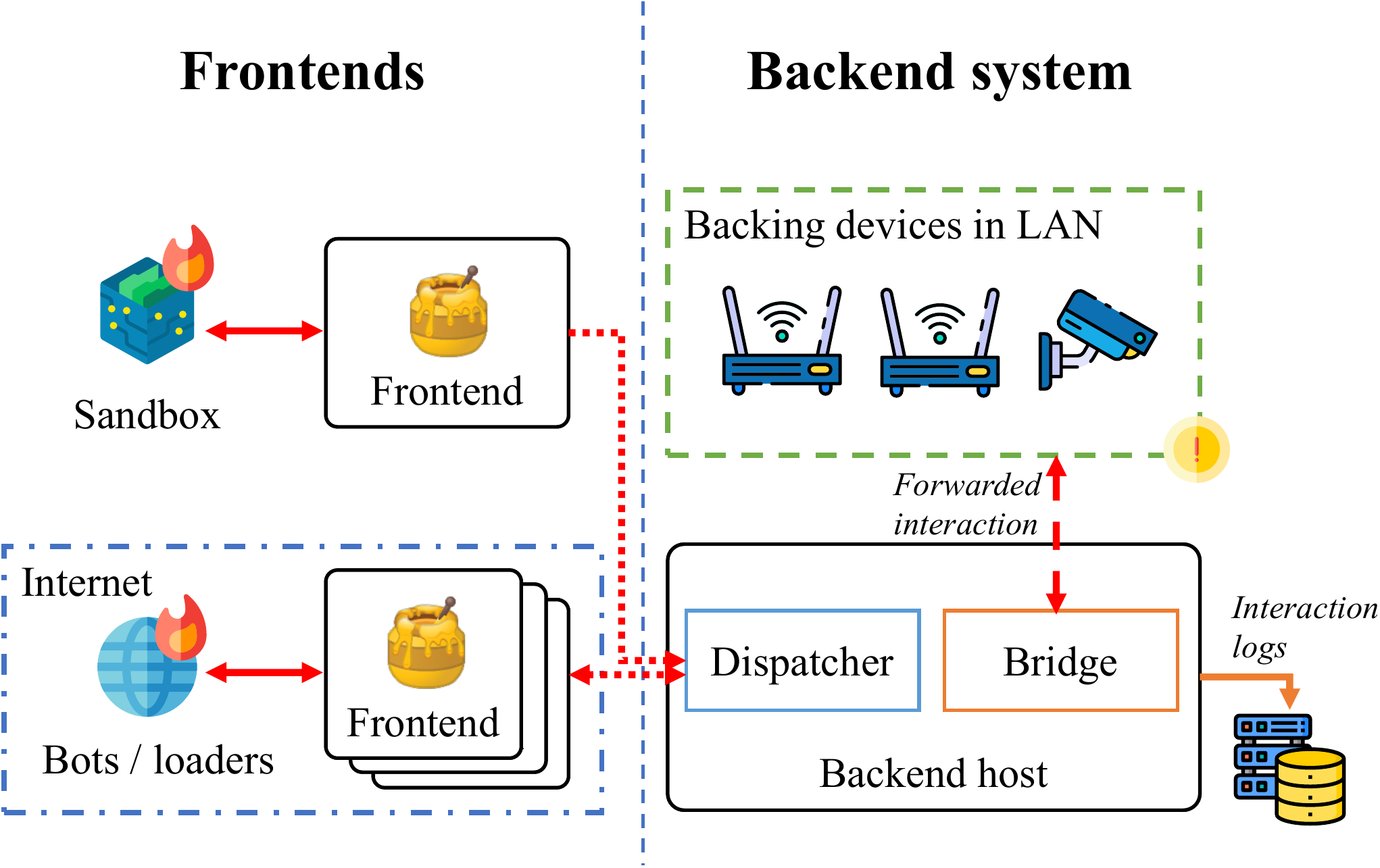}
  \caption{Honeypot system architecture.}
  \label{fig_sysarch}
\end{figure}


Weak credentials on telnet servers are the most common vulnerability among IoT devices, thus telnet-based intrusion is widely adopted as the primary attack vector by most botnets.
Analyzing and comparing the behavior of telnet loaders can be the most efficient way to make comparative and systematic studies on bot loaders.
In this study, we implemented a simple high interaction honeypot system to capture telnet interactions on the transport layer.
High interaction honeypot (HIH)\cite{guarnizoSIPHONScalableHighInteraction2017,wagenerHelizaTalkingDirty2011,vetterlHonwareVirtualHoneypot2019} handles a bunch of actual devices or services to show vulnerabilities and interact with attackers, which is opposite to \emph{low interaction honeypot} whose decoy is totally virtual.
Like many other honeypots in past studies, our honeypot consists of two major parts, the frontend, and the backend.
Frontends are deployed on cloud servers to interact with scanners and loaders directly, while the backend is deployed on a central server to dispatch requests to backing devices on a local network bridge.
The honeypot captures all interaction messages on the transport layer and generates conversation logs for the following analysis.
The overall architecture is depicted in Fig. \ref{fig_sysarch}.

The bridge module is a transparent proxy working on TCP protocol between the attacker and backing IoT devices.
It simply forwards bytes in telnet conversations to backing devices listed in Table \ref{tab_device}, but substitutes credentials to allow logging into the device.
The dispatcher will automatically pick a device for every telnet conversation.

\begin{table}
    \caption{Deployed backing devices during the experiment}
    \centering
    \begin{tabular}{ccc}
    \hline
    \textbf{Type}                          & \textbf{Device name} & \textbf{Software version}     \\ \hline
    \multirow{3}{*}{\textbf{Smart router}} & Lenovo Y1S           & PandoraBox git-6fcbaa5        \\ 
                                           & Netgear R7800        & OpenWRT 21.02                 \\ 
                                           & Netgear R6300v2      & KoolShare Merlin              \\ \hline
    \textbf{IP Camera}                     & Hikvision            & (Stock)                       \\ \hline
    \textbf{ONU}                           & CMCC I-120EM         & (Stock)                       \\ \hline
    \textbf{Other}                         & Raspberry Pi 3B      & Raspberry Pi OS Lite Jan 2021 \\ 
    \hline
    \end{tabular}
    \label{tab_device}
\end{table}

\label{subsec_ctrl_group}
In the final dataset, we expect a control group that all the samples generated by loaders or scanners from certain families, to help us evaluate the effectiveness and accuracy of our work.
Here we run a Hajime bot and a Mirai loader in a QEMU ARM sandbox and redirect requests on TCP/23 to a local honeypot.
The Hajime bot sample is provided by MalwareBazaar\footnote{https://bazaar.abuse.ch/}.
These conversation logs collected from the local sandbox are distinctly marked as the control group in the database.

Telnet is a text-based protocol commonly used for accessing a remote shell.
Telnet loaders leverage a set of command-line toolkits to start an automatic intrusion and set up a botnet client on victim devices.
Botmasters can take a codebase off-the-shelf (e.g. Mirai's loader) or implement new ones by themselves.
Also, they can modify the code to improve an existing loader or take some modules from others to enhance their own toolkit.
All these differences and relationships in their codebases can be reflected by the behavior fingerprints of loaders.
In the following sections, we leverage several common NLP methods to analyze the collected dataset to investigate these characters.

\subsection{Request Filtering and Concatenating}

Telnet, and its underlying TCP protocol, transfer messages in duplex mode.
Conversation logs from the honeypot consist of both request and response messages sorted by the time sequence.
However, in this work, we only focus on investigating loaders, so all the response messages in conversation logs are dropped in this step.
We also filtered out messages for credential prompt and protocol setup.
All remained requests are concatenated into a single ``request log" to represent all the behavior of a loader in the conversation, which is depicted in Fig. \ref{fig_tokenize}.

\subsection{Tokenize and N-gram vectorize}
\label{subsec_tok_vec}

\begin{figure}
  \centering
  \includegraphics[width=0.8\linewidth]{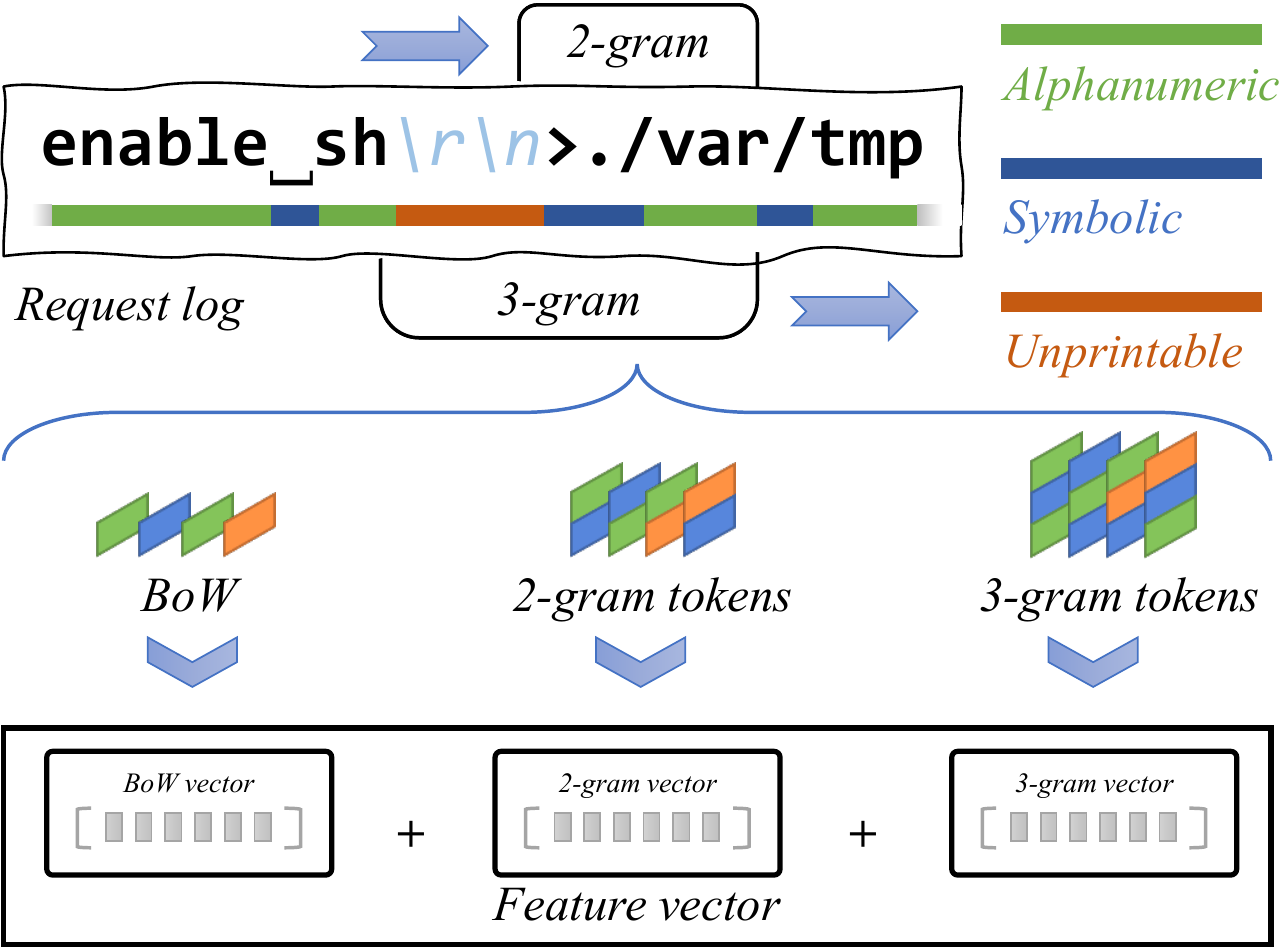}
  \caption{Byte-based tokenization and n-gram vectorization of request logs.}
  \label{fig_tokenize}
\end{figure}

In this step, we split request logs into tokens and generate n-gram feature vectors.
The process is depicted in Fig. \ref{fig_tokenize}.

To tokenize messages from text-based protocols, PRISMA\cite{kruegerLearningStatefulModels2012} uses a set of characters as delimiters.
However, in telnet messages, delimiters vary in different contexts. Diverse delimiters also reflect the syntax of a message, which could not simply drop.
The tokenize algorithm should preserve all bytes to retain features of message contents.
To satisfy the above needs, we tokenize request logs based on the type of bytes.
According to our early experiments, plain text is the major form of messages in a shell-based telnet conversation, while symbols and spaces compose semantic structures of shell commands.
Unprintable bytes are also valid payloads for telnet messages, which could be used for controlling the terminal and transferring files.
Based on this observation, we distinguish bytes into 3 types: alphanumeric, symbolic (including punctuations and spaces), and unprintable bytes.
Our method scans the type of every byte in a request log.
When finding two contiguous bytes of different types, our method cut the request log at this position.
Our method iterate on every byte of a request log to get the token sequence.
This keeps a token consisting of only one type of bytes and reserves all features from both printable and unprintable parts.

Our vectorize method uses a joint vector of BoW, 2-gram, and 3-gram to represent the semantic context of every token.
In shell command lines, the context of a token implies semantic information.
The bag-of-words (BoW) model only describes the counts of tokens, but the feature vector should reflect the previous and next tokens of them, which could be better described by the n-gram method.
The N-gram method slides an n-sized window on the sample from the beginning to the ending, generates an n-gram token, and finally generates the feature vector based on the count of these contiguous subsequences.
Therefore we represent the combinations of nearby tokens by 2-gram tokens.
Botmasters often make minor changes to intrusion toolkits.
In common situations, these variable tokens are surrounded by stable tokens constructing a context skeleton.
Thus we use 3-gram vectors to represent these minor modifications.

\subsection{Agglomerative Clustering}

To help define loader families based on their behaviors and further investigate their functions, we agglomerate request logs into clusters based on their similarities to partition the dataset.

Due to the deficiency of labeled request logs, unsupervised learning, especially clustering, is the optimal choice for this task.
However, most of the clustering methods rely on pre-defined amount of clusters to initialize themselves, which is ambiguous indeed for the task of family definition.
Moreover, our method should retain the pairwise similarity of clusters for further homological analysis.
In this case, we leverage the agglomerative clustering algorithm\cite{chidanandagowdaAgglomerativeClusteringUsing1978} to generate an elastic cluster definition on the dataset.

Agglomerative clustering is a bottom-top approach to building hierarchy clusterings.
Each observation, which is the feature vector of a request log in our method, starts in its own cluster.
A pairs of clusters \(\{C_{x},C_{y}\}\) are merged as one in every iteration, which is selected by finding the minumun inter-cluster distance \(\mathcal{D}\left(C_x,C_y\right)\) based on two hyperparameters: linkage criterion \(crit\) and distance metric \(d\).

\begin{equation}
  \begin{split}
  \mathcal{D}\left(C_x,C_y\right)=crit\left(d\left(x,y\right) \mid  x \in C_x,y \in C_y\right)
\end{split}
\end{equation}

These two hyperparameters can dramatically impact the shape of clusters.
Our method should select appropriate hyperparameters to evaluate the similarity of two request logs or clusters.

In agglomerative clustering, the distance metric \(d\) calculates the distance between two observations.
The most common metrics are the Euclidean distance and the cosine distance.
In this task, the Euclidean distance is the optimal choice for calculating the dissimilarity of request logs, which robustifies our method.
According to our previous experiments, the Euclidean distance performs better than the cosine distance in classifying request logs when there are shared components.



When two clusters to be evaluated consist of multiple observations, the linkage criterion determines the inter-cluster distance based on pairwise distances of observations.
While the average, complete and minimum linkage are common ones, we choose the \code{ward} as the linkage criterion, which minimizes the variance of the clusters being merged.

Finally, all the request logs can be hierarchically organized into an agglomerative binary tree, on which leaves correspond to request logs, and trunk nodes are agglomerative clusters.

The agglomerative clusters on the tree can be denoted as \(C\in\mathbb{N}\), where \(\mathbb{N}\) denotes the full set of them.
Every cluster \(C\in\mathbb{N}\) can be further splitted into two sub-clusters \(\{C_{1},C_{2}\}\) or merged into a super-cluster \(C^S\).
Cutting the tree at a given height \(\mathcal{T}\) by evaluating the condition \(F\left(C\right)\) on every \(C\in\mathbb{N}\) will produce a preliminary partitioning \(\mathbb{P}\subset\mathbb{N}\) at a selected precision:

\begin{equation}
  \begin{split}
  \mathbb{P}=&\{C \ | \ C\in \mathbb{N} \wedge F\left(C\right)\}\\
  F\left(C\right)=& \left[\mathcal{D}_{sub}\left(C^S\right) \geqslant  \mathcal{T}\right]  \wedge  \left[\mathcal{D}_{sub}\left(C\right) < \mathcal{T}\right]
\end{split}
\end{equation}

Note that we define the intra-cluster distance \(\mathcal{D}_{sub}\left(C\right)\) as the inter-cluster distance between its sub-clusters \(\mathcal{D}\left(C_1,C_2\right)\), which is the height of corresponding node on the agglomerative tree.
We leverage the implementation from Scikit-learn\cite{ScikitlearnMachineLearning} for the experiment.

\subsection{Template Extraction}
\label{subsec_template}

To identify the shared codebase of cousin loaders out of their request logs, we should make a comparative analysis to generate ``log templates'' for agglomerative clusters.

The agglomerative tree reveals the pairwise similarities of request logs and clusters.
Similar ones are coherent on the tree as the heterogeneous ones are far from each other.
So it is feasible to find a representative subsequence for every cluster based on the request logs by scanning the agglomerative tree from the bottom to the top.
We apply the Smith-Waterman algorithm\cite{smithIdentificationCommonMolecular1981} on every cluster to generate ``templates'' for them.

\begin{figure}
  \centering
  \includegraphics[width=0.8\linewidth]{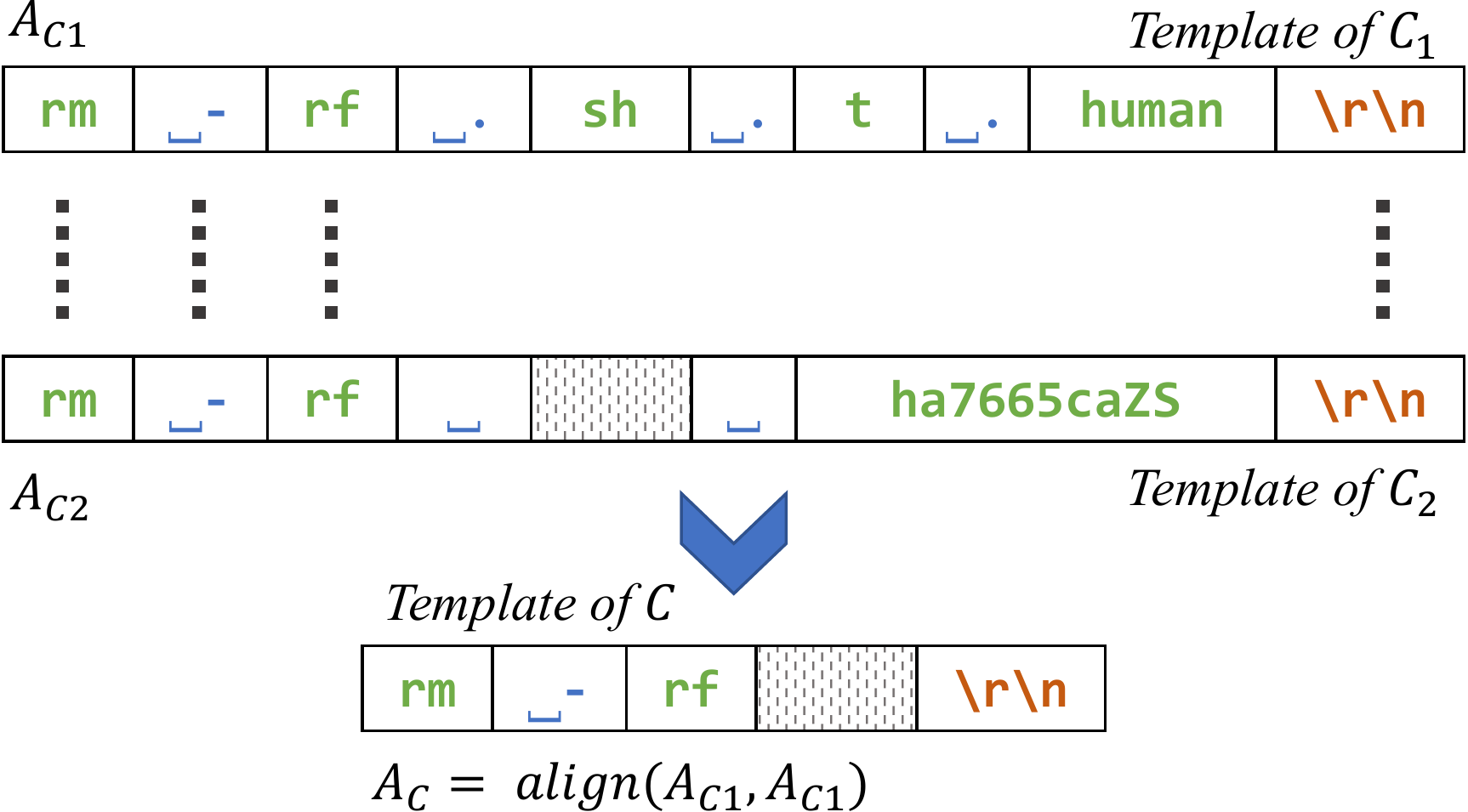}
  \caption{Smith-Waterman algorithm on the agglomerative tree.}
  \label{fig_smith_waterman}
\end{figure}

In our method, we use the tokenized sequence in the section \ref{subsec_tok_vec} to align two request logs.
As depicted in Fig. \ref{fig_smith_waterman}, the \(align(A_{C1}, A_{C2})\) function leverages the Smith-Waterman algorithm to scan both two tokens sequences from head to tail.
It performs sequence alignment on a local scope to determine similar regions between two strings.
Identical tokens at the same position will be aligned in the right position.
Our method adds placeholders (shadow cells in Fig. \ref{fig_smith_waterman}) to replace different tokens and help identical ones align, which finally generates a template.
For any cluster \(C\in\mathbb{N}\), the corresponding template \(A_{C}\) is generated recursively based on the templates of its sub-clusters \(A_{C1}\) and \(A_{C2}\).
Finally, every node on the agglomerative tree will get a representative ``template'' which could be easily interpreted.

\subsection{Clustering Refinement}

However, although we empirically cut the agglomerative tree to roughly classify the request logs, this is far from being taken as the final class definition. The unique \(\mathcal{T}\) value may not fit all branches on the tree, so we have to interpret the templates of clusters in \(\mathbb{P}\) and adjust the partitioning to calibrate the family definition.
Here we empirically set some criteria for accepting or denying a merged cluster:

\label{manual_criteria}
\begin{itemize}
  \item Commands is the major evidence for the judgment, their arguments and calling order should also be taken into consideration.
  \item When there are complex statements in command lines, the structure and syntax are more important than its component commands.
  \item Tokens used for manifesting botmasters should be ignored unless it appears with new commands or different arguments.
\end{itemize}

We start from nodes in \(\mathbb{P}\) and check corresponding templates to decide if a cluster should be kept, merged, or further split.
Finally, for every host we observed, we collect all related samples and their class labels, then pick the label with the maximum count as the label of this loader host.

\section{Evaluation}
\label{sec_eval}
In this section, we discuss functions and behaviors of active loaders based on the aforementioned methods and make a conclusion about their homologies.

\begin{figure}[tpb]
  \centering
  \includegraphics[width=\linewidth]{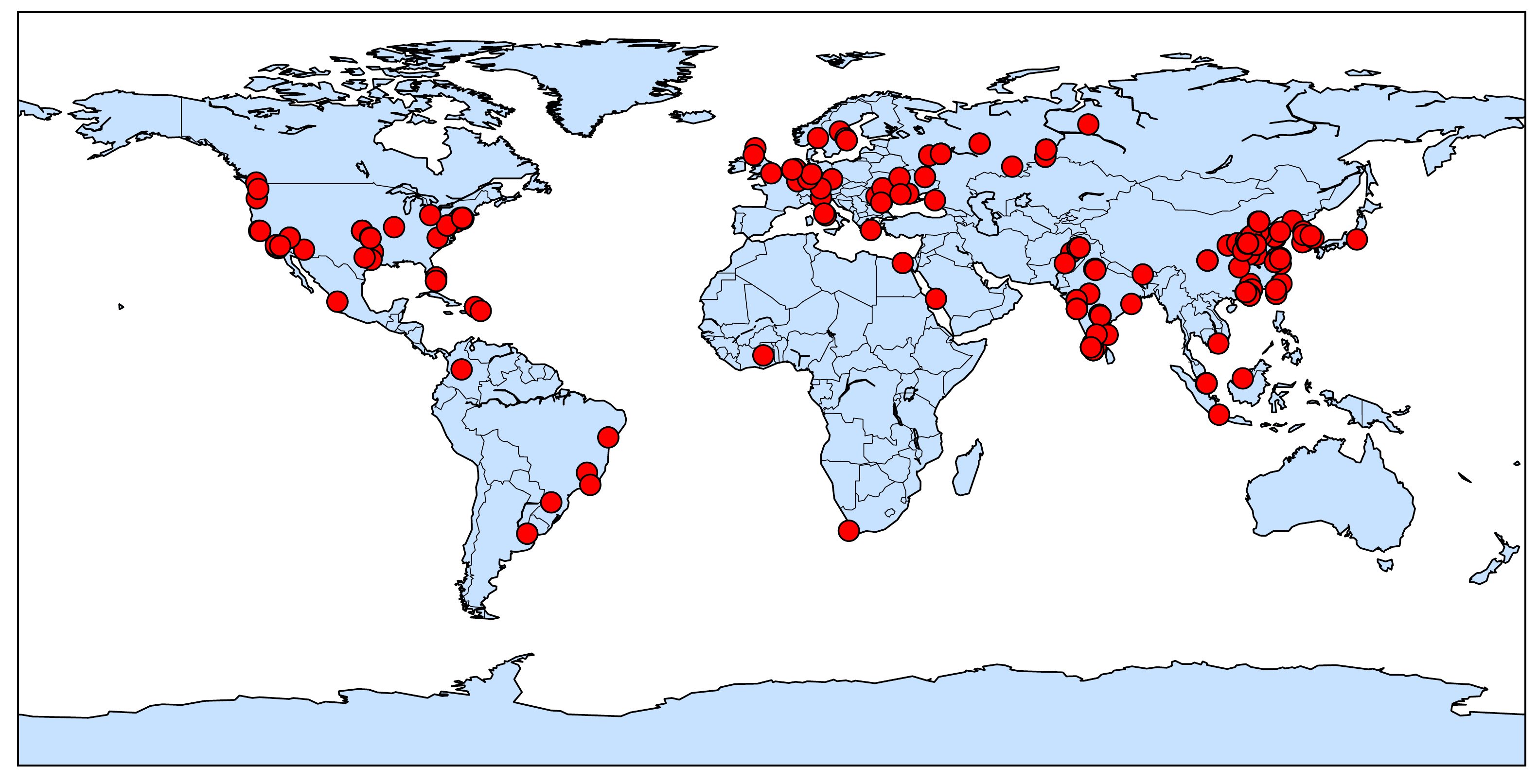}
  \caption{Geographical distribution of loader hosts.}
  \label{fig_geo_dist_map}
\end{figure}

\subsection{Captured Dataset}

The following analyses is based on conversation logs we captured from November 14 to December 31 in 2021.
We observed that 527 hosts had interacted with our honeypot system.
These hosts distributed across 35 countries/regions and 121 autonomy systems as listed in Fig. \ref{fig_geo_dist_map} and Table \ref{table_as_dist}.
To reduce the scale of the dataset, we took less than 20 conversation logs for each host and selected 4,855 out of over 3 million items.
This produced 481 unique request logs and generated feature vectors of 8,083 dimensions.

\begin{table}[htbp]
    \centering
    \caption{Distribution of loader hosts in autonomy systems}
    \begin{tabular}{cccc}
      \hline
      \textbf{Country} & \textbf{Name}                                     & \textbf{Hosts} & \textbf{Logs} \\ \hline
      China            & CHINA UNICOM China169 Backbone                    & 184            & 2145              \\
      India            & Hathway IP Over Cable Internet                    & 29             & 361               \\
      Russia           & MTS PJSC                                          & 29             & 360               \\
      China            & Chinanet                                          & 28             & 286               \\
      India            & National Internet Backbone                        & 20             & 243               \\
      \hline
      \textit{Others}  & \textit{}                                         & \textit{237}   & \textit{1454}     \\
      \hline
    \end{tabular}
    \label{table_as_dist}
\end{table}

We made a simple statistical analysis of loader hosts' geographical distribution.
It shows that China, Russia, India, and the United States are preferred countries for botmasters to setup loader servers.
Over 75\% loaders are located in these 4 countries.
While it is common sense for Mirai-based botnets to build infrastructures on public cloud servers, we can still see there are lots of servers deployed in consumer networks like ASN4837 and ASN17488.
This means that some botnets might have a different architecture from Mirai, and loaders are also distributed across multiple networks and regions.

\subsection{Family Definition}

In this step, we leverage agglomerative clustering and try to define several families for bot loaders based on the collected dataset.

\subsubsection{Clustering overview}

\begin{figure}[tpb]
  \centering
  \includegraphics[width=\linewidth]{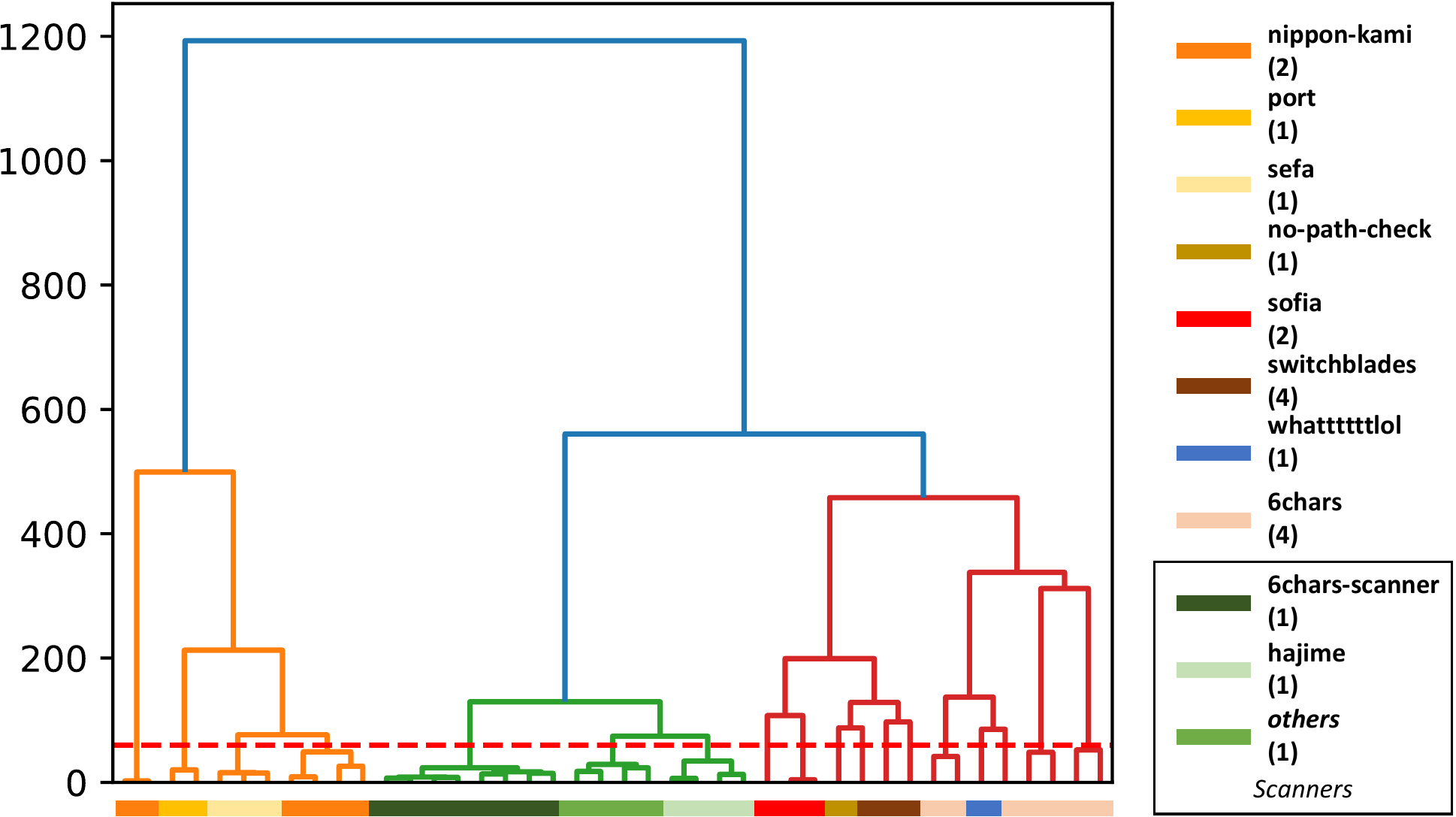}
  \caption{Dendrogram of clustering result and family definitions. The colored bars indicate the family definition of the agglomerative tree. The counts of member clusters of each family are listed in the legend on the right side. Red horizontal line for \begin{math}\mathcal{T}=60\end{math}. }
  \label{fig_dendrogram}
\end{figure}

The agglomerative clustering algorithm generated a tree with a height of 1193.07, which dendrogram is depicted in \ref{fig_dendrogram}.
While the tree is relatively tall, most of the branches are at a height below 200.
Minority branches at a higher height may manifest significant differences in corresponding bot loaders' behaviors.
Based on the method in Section \ref{subsec_template}, we recursively generated templates for 480 non-singleton clusters to describe their common behaviors.

\subsubsection{Threshold selection}

Although the agglomerative clustering algorithm hierarchically clustered all the samples we captured into a tree, we should cut the tree at a threshold \begin{math}\mathcal{T}\end{math} to obtain an elementary partition on the dataset for further manual analyses.
Griffioen and Doerr \cite{griffioenExaminingMiraiBattle2020} examined the battle among IoT botnets. Although this study listed 39 families according to their observation, most of the behavior analyses only focused on 14 botnets which are the most active ones.
In Wang's text-based analysis on botnet reports\cite{wangEvolutionaryStudyIoT2021}, the experiment shows that despite there is a large number of observed botnets, most of them are actually minor revisions or just aliases of some popular botnets.
Combining conclusions from these studies, we expect a \begin{math}\mathcal{T}\end{math} value producing around 20 clusters, which is a reasonable amount for covering active bot loaders.
The limited number of clustering also makes it convenient to examine their templates and further refine the family definition.
Based on the clustering result, we empirically set \begin{math}\mathcal{T}=60\end{math}, which is displayed as a red horizontal line in the Fig. \ref{fig_dendrogram}.
This produced 19 clusters from the previous clustering result.

\subsubsection{Interpreting the agglomerative tree}

\begin{table*}
  \centering
  \caption{Index Table of Loader Functions Discussed in Section \ref{subsec_behavior}}
  \includegraphics[width=0.9\textwidth]{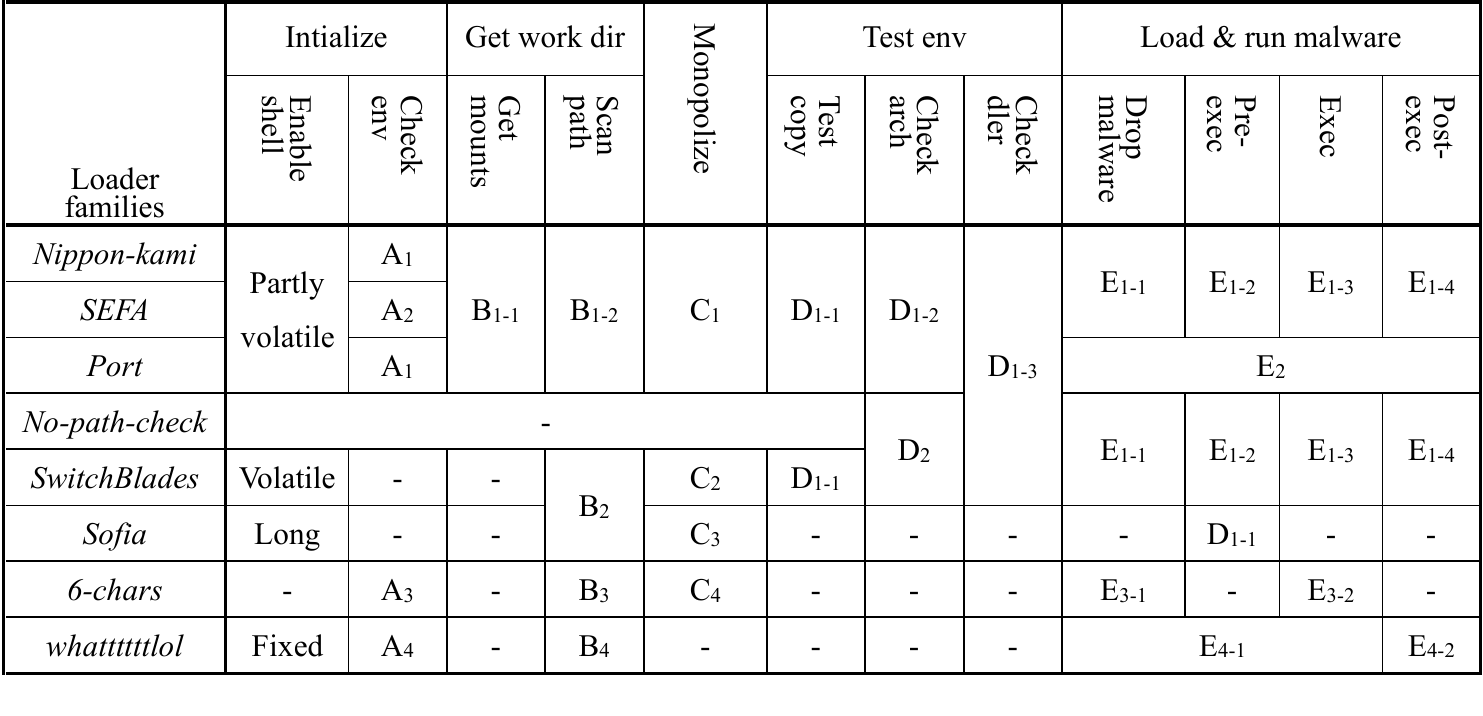}
  \label{fig_table_func}
\end{table*}

Based on the extracted templates, we evaluated these clusters based on the aforementioned criteria in section \ref{manual_criteria}, and finally figured out several families out of the dataset.
We traverse their sub-clusters and find some representative token as their name, which may not follow the common rule of naming a botnet.
We listed these families and indexed their functions in Table \ref{fig_table_func}.

The yellow branch in the Fig. \ref{fig_dendrogram} consists of 3 families.
These families share the original Mirai codebase and made minor revisions to fit their botmasters' needs:

\begin{itemize}
  \item \emph{Nippon-kami.} This is the family which directly derived from Anna-Senpai's release\cite{anna-senpaiFREEWorldLargest2016}. ``Query token'' which queries the status of command are frequently modified to declare the botmaster's identity, and the initialize command list to enable shell interface is the most volatile part all over the class. 
  \item \emph{SEFA.} This loader manifests itself with the token ``SEFA'' in initial commands and the query token, but obviously share the same codebase with the \emph{``nippon-kami''} family.
  \item \emph{Port.} This loader calls openssl for some unknown reason and we have not seen it dropping any malware yet. We observed this loader only in late December 2021, which means it might have retired in January 2022.
\end{itemize}

We also found several families that were located relatively further to the yellow branch, so we color the branch red in Fig. \ref{fig_dendrogram}.
These 5 families are implemented significantly different and modified some crucial components to satisfy the needs of their botmasters:

\begin{itemize}
  \item \emph{No-path-check.} This is a tailored variant of the \emph{``nippon-kami''} loader. All the commands prior to the step of checking architecture are removed and it simply uses the default working directory of the logged-in user.
  \item \emph{SwitchBlades.} This loader derived the framework of \emph{nippon-kami}, but has a different implementation to detect writable directory, which actually acts like \emph{Sofia}. We took the name from the first sample we saw in this family, but the identification token is volatile and we also see others including \code{Layer1} and \code{skull}.
  \item \emph{Sofia.} This loader extremely extended the initial command list. It implemented a different method to detect writable directories and use a hard-coded list to scan the file system.
  \item \emph{6-chars.} The most significant feature of it is the query string. Unlike other loaders, loaders of this family generate 6 random escaped characters as a query string for every session. We noticed that some scanners also share the same pattern, but here we distinguish them as an independent \emph{6-chars-scanner} family and won't do functional analysis on them.
  \item \emph{``whattttttlol''.} This loader doesn't have any shared parts with the aforementioned loaders, which makes its behavior significantly different from others. It runs a fixed command list and downloads multiple shell scripts named ``\code{whattttttlol*.sh}'' to load the malware.
\end{itemize}

\subsubsection{Clustering of scanners}
During the experiment, we notice that some samples in particular clusters do not carry any downloading commands.
These data all distributed on the green branch in Fig. \ref{fig_dendrogram} but can be divided into 3 clusters.
The data of previously mentioned \emph{6-chars-scanner} also distributes on this branch.
Based on this phenomenon, we regard them as scanners (named ``others'') which simply probe exploitable hosts but never load an executable.

\subsubsection{The control group}
Our method successfully classified the control group dataset we mentioned in Section \ref{subsec_ctrl_group}.
The request logs generated by the local Hajime scanner are located closely with other scanners and Anna-senpai's Mirai loader is located along with its variants in the yellow branch.
This result substantiated the effectiveness of our method in clustering similar loaders and distinguishing different malicious adversaries.

\subsection{Behavior Investigation}
\label{subsec_behavior}
\def\subscr#1#2{#1\textsubscript{#2}}

\begin{figure*}[htbp]
  \centering
  \includegraphics[width=0.9\textwidth]{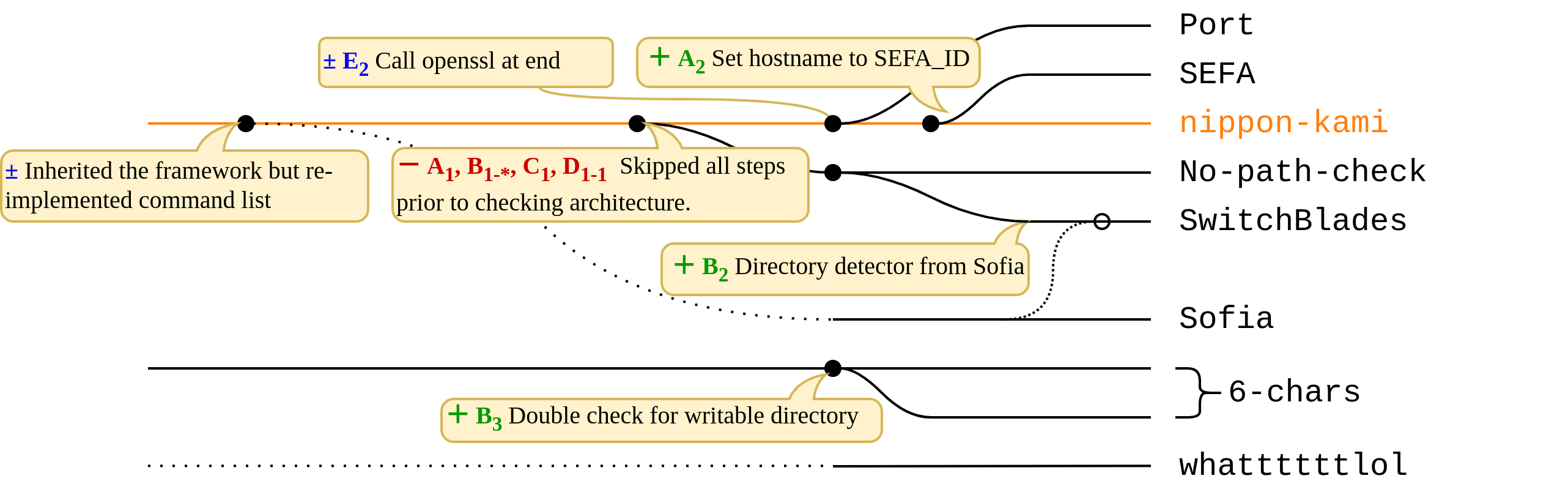}
  \caption{Homology dendrogram of captured loaders. Horizon lines depict identified families, while the yellow one highlights the \emph{nippon-kami} released by Anna-Senpai. In the description of a connecting line between two families, green ``+'' indicates adding functions, red ``-'' indicates removing functions, and blue ``\textpm'' indicates modifications. All the functions are denoted by indexes in Table \ref{fig_table_func}.}
  \label{fig_dendro_manual}
\end{figure*}

\lstset{
basicstyle=\small\ttfamily,
columns=flexible,
breaklines=true,
frame=single
}

\begin{figure}[bp]
  \emph{\subscr B{1-2}}
  \begin{lstlisting}
busybox echo -e '\\x6b\\x61\\x6d\\x69/proc' > /proc/.nippon; 
busybox cat /proc/.nippon; 
busybox rm /proc/.nippon
  \end{lstlisting}
  \emph{\subscr B{2} and \subscr B{3}}
  \begin{lstlisting}
>/var/tmp/.file && cd /var/tmp/
  \end{lstlisting}
  \label{codefig_dirprobe}
  \caption{Sample codes of the ``Get working directory'' function denoted by indexes in Table \ref{fig_table_func}.}
\end{figure}

\begin{figure}[bp]
  \emph{\subscr D{1-1}}
  \begin{lstlisting}
/bin/busybox cp /bin/echo sefaexecbi; >sefaexecbi; /bin/busybox chmod 777 sefaexecbi; 
  \end{lstlisting}
  \emph{\subscr D{1-2}}
  \begin{lstlisting}
/bin/busybox cat /bin/echo
  \end{lstlisting}
  \emph{\subscr D{2}}
  \begin{lstlisting}
/bin/busybox cat /bin/busybox || while read i; do echo $i; done < /bin/busybox
  \end{lstlisting}
  \caption{Sample codes of the ``Test environment'' function denoted by indexes in Table \ref{fig_table_func}.}
  \label{codefig_envprob}
\end{figure}

In this section, we interpret the Table \ref{fig_table_func} vertically to make a comprehensive comparison of functions based on extracted templates and make a conclusion about their shared codebases.
In this section, we denote all the discussed components by alphanumeric indexes in Table \ref{fig_table_func}.
By discussing all the relationships of their components, we draw a dendrogram (Fig. \ref{fig_dendro_manual}) to show the homology of functions among all classes we defined.
The yellow line represents nippon-kami which is directly derived from Anna-Senpai's release.
The intrusion process can be distinguished into 6 major functions:

\subsubsection{Initialize} 

To enable the full shell on an IoT device in the telnet conversation, loaders execute an initialization command list to cover all the possible commands.
Then, the loader checks the environment to scan existing bot instances or security services.

As Nippon-kami, SEFA, and Port share the same codebase, their initial command list varies slightly depending on their botmasters.
Then they run \code{ps} command to check suspicious processes in the environment (\subscr A{1}).
The SEFA loader added a command for modifying the victim's hostname to \code{SEFA\_ID:\textit{<4-digit numbers>}} (\subscr A{2}) which could be an identification of bots in the botnet.
Sofia has an extended list for initializing commands, but the commands checking environments have been removed.
While the whattttttlol holds a fixed command list, it runs \code{ls /home} to scan files in the directory (\subscr A{3}).
6-chars only checks the \code{wget} command here, which is commonly executed later in other loaders (\subscr A{4}).

\subsubsection{Get working directory}

Most of the loaders require a writable directory to temporarily drop the executable.
The directory scanner is an important component of loaders which has many different implementations.
Nippon-kami, SEFA, and Port get mounted directories (\subscr B{1-1}) and run a composed statement (\subscr B to check its write privilege in this step{1-2}).
It prints an escaped token \code{kami\textit{<path>}} using \code{echo} to a file named \code{.nippon}, reads the file and test if the file is successfully written.
Port uses a different escaped token and file name in this step.

SwitchBlades, Sofia, 6-chars and whattttttlol all use a hard-coded list instead of scanning the mounted filesystems at the run-time.
Both SwitchBlades (\subscr B{2}) and 6-chars (\subscr B{3}) run an ``\code{\&\&}'' (and) statement to test writable privilege using a hard-coded list, while the list is slightly different in these two families.
A variant of 6-chars runs this step twice, which shows a difference in the request logs.
SwitchBlades uses a return to join these statements.
The last usable directory will be used in the following steps.

Whattttttlol uses a simple ``\code{||}'' (or) statement to join multiple \code{cd \textit{<directory>}} (\subscr B{4}).
This changes the working directory to the first available one in the hard-coded list, regardless of its writable privilege.

Some of the implementations are displayed in Fig. \ref{codefig_dirprobe} to show their difference.

\subsubsection{Monopolize} 

It is extremely normal for vulnerable devices to be infected repeatedly.
The newcomer should kill the existing bot client to acquire control of the device.
Most loaders will try eliminating competitors by deleting certain files in a built-in list.
Nippon-kami and its variants use \code{.sh .t .human} (\subscr C{1}), Sofia use \code{.file .cowbot.bin retrieve cowffxxna} (\subscr C{3}), and 6-chars removes \code{.i} only (\subscr C{4}).
SwitchBlades tries to delete two files that may be related to its old executables. The list is unstable among different variations (\subscr C{2}).

\subsubsection{Test environment}

In this step, the loaders should probe the CPU architecture and available download commands to decide which executable will be loaded and how it will be loaded.
The classic method (\subscr D{1-1}, \subscr D{1-2}, \subscr D{1-3}) tests \code{cp} command, prints \code{/bin/echo}, then test \code{wget} and \code{tftp} commands.
The CPU architecture can be obtained by parsing an existing executable on the device.
\subscr D{1-2} implement this by running \code{cat /bin/echo}.
For \subscr D{2}, no-path-check and Sofia prints \code{/bin/busybox} to obtain the same information.
In case of \code{cat} command is unavailable, they also use a shell-based \code{while read} statement to print the file.
For cross-platform botnets, this will decide which cross-compiled executable will be dropped in the next step.

Some of the implementations are displayed in Fig. \ref{codefig_envprob} to show their difference.

\subsubsection{Drop \& run malware} 

Loader download the executable, setup the environment, execute it and clean all its trails in the final stage.
In this step, loaders \code{cd} to the probed working directory leverage the available download tool detected in the previous step to drop the malware.

Specifically for nippon-kami (\subscr E{1-1}), if both \code{wget} and \code{tftp} are unavailable, it will run the fallback command and load the whole file with \code{echo} command and a stdout redirect statement.
SEFA, no-path-check and SwitchBlades all reused this implementation (\subscr E{1-*}).

6-chars leverages an ``\code{||}'' (or) statement to call multiple commands sequentially (\subscr E{3-*}) until a command succeeds.
Whattttttlol called multiple commands sequentially to download and run 4 scripts and deleted them all after execution (\subscr E{4-*}).

In our collected dataset, we did not observe Port and Sofia downloading executables in this step.
Instead, Port called \code{openssl} for unknown reason (\subscr E{2}), while Sofia only checked writable privilege in the current directory (\subscr D{1-1}).

\subsubsection{Query token}

This command is designed to run after every command to identify this step has finished.
The loader would wait for a specified token in response before it goes on.

\begin{figure}[tbp]
  \centering
  \includegraphics[width=0.9\linewidth]{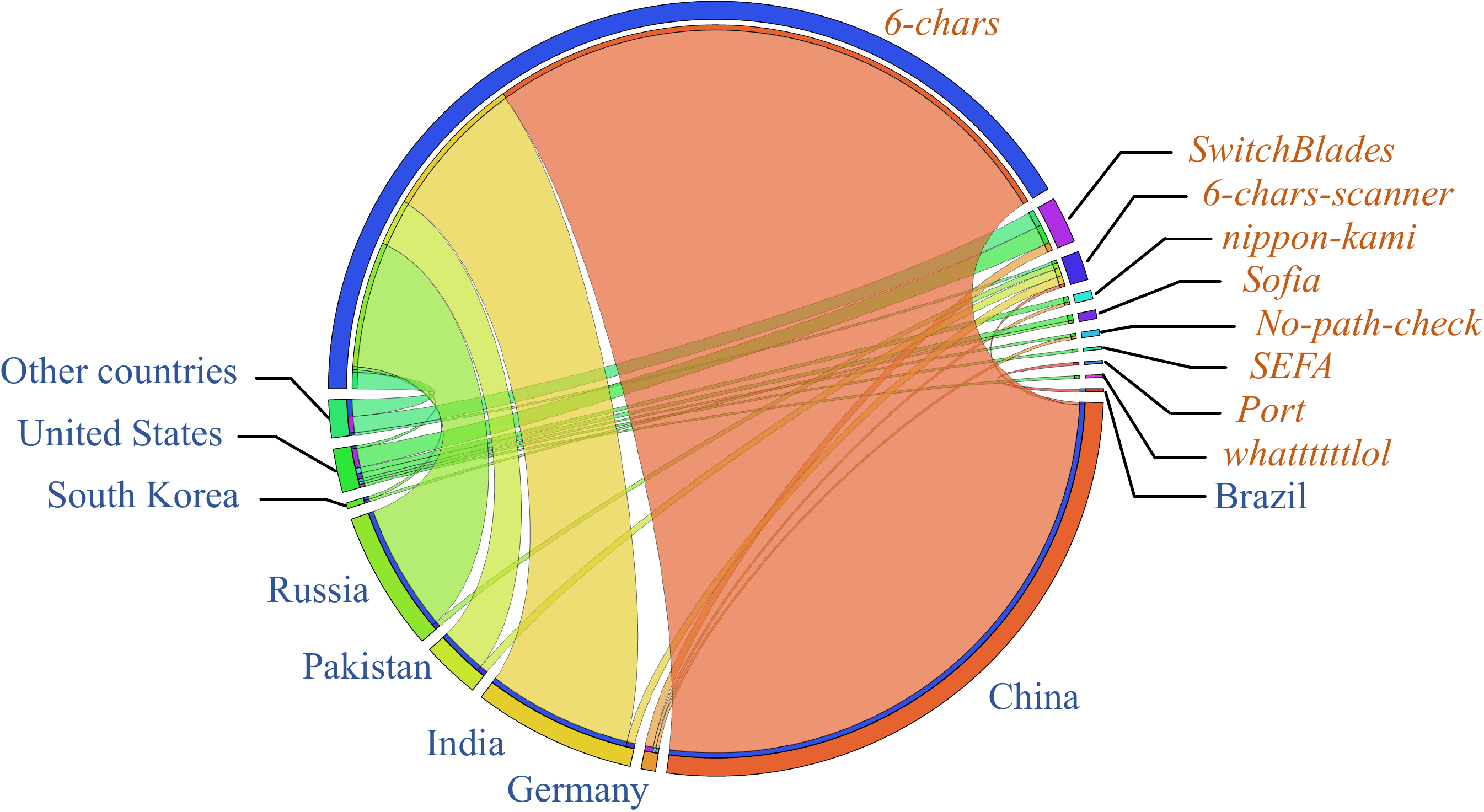}
  \caption{Geological distribution of observed loader families.}
  \label{fig_family_geo_dist}
\end{figure}

\subsection{Geological Distribution of Loader Families}

We draw a circos graph to further demonstrate the geological distribution of different loaders.
Scanners except 6-char-scanner are dropped.
As shown in Fig. \ref{fig_family_geo_dist}, 6-char loaders are distributed diversely across China, India, Russia, and Pakistan, while nippon-kami and all its variants are only observed to have a few loader hosts distributed in a small number of countries.
This inspired us to reach the following two conclusions.

\begin{itemize}
    \item
      While Mirai is designed to have a centralized loader system, many other botnets also try integrating loader programs into bot clients, including the botnet using 6-chars as the loader.
      A distributed loading system makes the botnet more robust when coping with taking-downs on single loader servers.
      Hajime had already implemented this leveraging its modularized design.
    \item
      For small cybercrime teams, developing on Mirai's codebase is still a good choice.
      The existence of SEFA and Port is a piece of important evidence on this.
      Some other teams also implement their loader for some reason.
      These small botnets implemented some new features but do not have the redundant resources to build multiple loader servers.
      We conclude that there is still a long way to go before Mirai disappears if ever possible.
\end{itemize}

\section{Discussion}
\label{sec_discussion}
For this study, we focus on investigating loaders based on the interaction logs.
However, this limited methodology is far from ultimate for mining all covert information of botnets.

\subsubsection*{Associated malware}

Bot client is the most volatile part of botnets.
Some existing research made lineage analyses on executables captured by honeypots.
However, our work lacks aspects on bot clients.
Combining these two analyses to analyze relationships among loaders and clients may extend the conclusion of homology analysis, which is a feasible and promising field for botnet studies.

\subsubsection*{Family definition}

Our work draws a conclusion based on the result of agglomerative clustering.
The empirical value of \begin{math}\mathcal{T}\end{math} determined the family definition of bot loaders.
However, due to the heterogeneity of loaders' behavior, a homogeneous \begin{math}\mathcal{T}\end{math} cannot fit all situations.
For example, We found that whattttttlol was misclassified along with 6-chars, which reflects our method is still deficient to make a clear definition of classes.
The cutting condition of the agglomerative tree should be further discussed in the following studies.
Note that the query tokens and credentials are not fully investigated, which could miss critical information of loaders.

\subsubsection*{Early stage detection}

For IDS systems, the loading stage is one of the best chances to intercept emerging attacks.
Our methodology for analyzing network messages push the identification further to recognize the identity of attackers, which is of great importance in reinforcing the security of the Internet.

\section{Conclusion}
\label{sec_conclusion}
In recent years, the open-source botnets spawned a mass of variants, which is a major threat to IoT devices.
While understanding the behaviors of botnet malware is crucial, their infrastructures, especially loader servers, are not systematically investigated yet.
In this work, we proposed a novel aspect of homology analysis on bot loaders.
We analyze loaders' implementation using honeypot logs and evaluate their relationship by shared components.
Our experiment consists of 3 components.
First, we collect interaction logs using a high interaction honeypot system to represent the functions and behaviors of loader hosts.
Second, we use agglomerative clustering to cluster these logs, fit all relationships onto a tree, and cut at a threshold to classify the loaders.
Finally, we evaluate the tree bottom-top and leverage the align algorithm to evaluate shared components in each loader family.
We find that the ecosystem of Mirai is still active despite numerous takedowns.
New variants are created to satisfy botmasters' needs in many aspects.
We also discover that some small-scale botnets come and fade away in a very short timespan.
The security community should further develop the methodology of botnet investigation to enhance the knowledge of botnets' relationships, which could help eliminate cybercrimes more efficiently.

\bibliographystyle{IEEEtran}
\bibliography{IEEEabrv, lib}

\end{document}